\documentclass[10pt]{iopart}
\usepackage{graphicx}  
\usepackage{times}  

\begin{document}
\renewcommand{\textheight}{7.48truein}

\newcommand{\Decagon}{{13\AA{}~ Decagon}}
\newcommand{\Decagons}{{13\AA{}~ Decagons}}
\newcommand{\Dec}{\Decagon}
\newcommand{\Decs}{\Decagons}
\newcommand{\Star}{{Star cluster}}
\newcommand{\Stars}{{Star clusters}}

\newcommand{\Mihet}{Mihalkovi\v{c} {\it et al}}
\newcommand{\Henet}{Henley {\it et al}}
\newcommand{\Guet}{Gu {\it et al}}
\newcommand{\et}{{\it et al}}
\newcommand{\rr}{{\bf r}}

\newcommand{\ubar}{{\bar u}}

\vskip 0.2 true in

\title
[Co-rich decagonal Al-Co-Ni] 
{Co-rich decagonal Al-Co-Ni: predicting structure, orientational
order, and puckering}

\author{\rm NAN GU\dag, C.~L.~HENLEY$^{*}$\dag,  
and M.~MIHALKOVI\v{C}\ddag }


\address{\dag\ Dept. of Physics, Cornell University,
Ithaca NY 14853-2501}

\address{\ddag
Institute of Physics, Slovak Academy of Sciences, 84228 Bratislava, Slovakia.}

\vskip 0.2 true in

\begin{abstract}
We apply systematic methods 
previously used by Mihalkovi\v{c} {\it et al.}
to predict the structure of the `basic' Co-rich modification of the 
decagonal Al$_{70}$Co$_{20}$Ni$_{10}$ layered quasicrystal, based on
known lattice constants and previously calculated pair potentials.
The modelling is based on Penrose tile decoration and uses 
Monte Carlo annealing to discover the dominant motifs, which are
converted into rules for another level of description.
The result is a network of edge-sharing large decagons on a
binary tiling of edge 10.5 \AA{}.  
A detailed analysis is given of the instability of a four-layer
structure towards $c$-doubling and puckering of the atoms out
of the layers, which is applied to explain the (pentagonal) orientational
order.
\end{abstract}



\vskip 0.2 true in

\section{Introduction}





A fundamental problem of theoretical solid-state chemistry is, 
{\it assuming} the ability to compute perfectly the total energy
of  a complex compound,
to discover which arrangement of the atoms minimizes that energy.
Quasicrystals offer a wonderful opportunity to face this
problem, as many different arrangements are low in energy, 
but this approximate degeneracy is resolved by more subtle
interactions, often in several steps of a hierarchy of 
length scales.   

Consider decagonal Al-Co-Ni, the pre-eminent
decagonal Al-TM quasicrystal (TM= transition metal), 
a phase which is divided into many modifications~\cite{Ritschet1998}, of which the
simplest are `basic' Ni-rich (around Al$_{70}$Co$_{10}$Ni$_{20}$) and 
`basic' Co-rich (around Al$_{70}$Co$_{20}$Ni$_{10}$).
\Mihet~\cite{Mihet2002} proposed a general method to discover the 
ground-state structure, 
taking as inputs the interatomic pair potentials,
derived by Moriarty and Widom~\cite{Moriarty1997} 
using `generalized pseudopotential theory',
and the experimentally determined lattice constants. 
One initially describes the quasicrystal by 
a Penrose tiling of edge $a_0 \approx 2.45$ \AA{} 
with properly placed candidate sites on the tiles. 
Monte Carlo annealing, in which atoms hop on these sites,
is used to find low-energy configurations. 
(This approach was used earlier by Cockayne and Widom~\cite{Cockayne1998}
\newpage \noindent
for decagonal Al-Co; \Mihet~\cite{Mihet2002} added tile rearrangements as
a helpful Monte Carlo move.). 
In the next stage, motifs abstracted from these results are converted into  
{\it rules}  to decorate  larger 
tiles, thereby reducing the degrees of freedom and permitting
larger simulation cells in which further orderings can be seen.

\section{Structure motifs and idealized decoration}
\label{sec:deco}

\begin{figure}
\includegraphics[width=4.6in,angle=0]{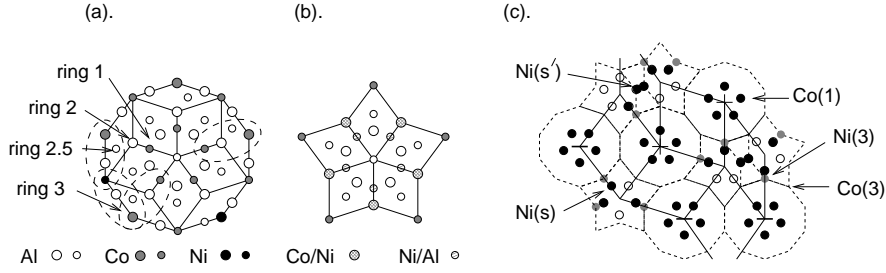}
\caption{Decoration model
(a). Decoration of a \Decagon, named in concentric rings.
Size of circles indicates which of the two layers;
some third-ring `Co' atoms become Ni (as shown) but
not many.
Dashed ovals mark four (out of eight) of the places
showing Co-Al-Al-Co in projection, which after relaxation
become a channel like figure~\ref{fig:channel}a.
(b). Decoration of \Star.  There are additional Al along
the outer edges.  
(c). \Decagons~ and \Stars~ shown with dashed edges, placed
on vertices of the binary-tiling
rhombi (edge $\tau^2 a_0 \approx 10.5$\AA{}).
Filled circles are TM atoms in \Dec~ and \Star~ interior.
The $\pm$ sign on each \Dec indicates $\sigma_i=\pm 1$.
Ni(3) sites are shown in grey; there are Co atoms (not shown)
on all other corners of the \Decs.}
\label{fig:clusters}
\end{figure}

We have applied this method to the `basic' Co-rich composition
(more complete accounts may be found in Refs.~\cite{Guet2005a,Guet2005b}.)
From the initial stage simulation
it became clear that the dominant motif is a
decagon with edge $\tau a_0 \approx 4$\AA{} and hence diameter
$2\tau^2 a_0 \approx 12.8$\AA{};
in this rest of this paper, the
term `13 \AA~Decagon' refers exclusively to this object.
The exact occupancies 
(fig.~\ref{fig:clusters}a) were
clarified in the second stage of simulations.
These were based on $\tau a_0$-edge
rhombi in arrangments favoured the densest possible packing of
non-overlapping 13\AA~decagons made from the rhombi.
(Overlaps were verified to be unfavourable energetically.)
The atoms in a cluster are grouped into concentric rings
(see figure~\ref{fig:clusters}a).  Just inside the perimeter,
a few Al sites form a partial `ring 2.5' with elusively
context-dependent occupancies.  (The Al positions on each \Dec~ edge (ring 3)
also may depend on the \Dec's environment.)
Almost all the TM atoms in the \Dec~ are Co, which is 
presumably why it emerges at higher Co content.  
(This was also discovered by Hiramatsu and Ishii~\cite{Hiramatsu2004}, 
who did an exploratory 
study using the same code of a wider range of AlCoNi compositions.)

The best arrangements for edge sharing of the \Decs~ amounts to a 
10.5~\AA{} edge `binary tiling'
\cite{Lanconet1986, Widomet1987},
with the \Decs~ sitting on the Large-atom sites of the binary tiling.
Filling the space between is a secondary motif, which is centred
on the Small-atom sites, which we call the `\Star' and show in
figure~\ref{fig:clusters}(b).
The centre of this cluster has a pentagon of five candidate
TM sites, which are variably occupied by TM (almost always Ni)
or Al.  

The TM occupancies show context dependence, which we will
approximate with a very simple rule (figure~\ref{fig:clusters}c). 
We have four classes of TM atom.
On each of the binary tiling's Large vertices 
we have a \Decagon~ with five Co(1) in its centre and ten TM(3)
on its perimeter (ring 3).
(For visual clarity, the TM(3) are not shown
in figure~\ref{fig:clusters}c.)
Now, on each (binary-tiling) edge 
connecting a \Decagon~ to a \Star~ centre, a \Star~ TM is
strongly favoured or disfavoured according
to whether the \Decagon's first ring has Al or TM on the
same edge.  (The distance between 
these sites is $5^{1/2} a_0 \approx 5.6$\AA{}, which is
a maximum of $V_{\rm TM-TM}(r)$ \cite{Moriarty1997}.]
So, on {\it every} binary-tiling rhombus 
exactly half the edges 
do not pass over a Co(1): let us place on each such 
edge a Ni$(s)$ within the \Star,
and make the TM(3) site a Ni(3); all other TM(3) become Co(3).
(As in the `basic' Ni-rich composition,
the difference between Ni and Co potentials
favours Ni occupancy in TM-TM pairs and Co in lone TM sites.)

Finally, as this too is energetically favoured, we place 
2 Ni$(s')$ wherever two \Stars~ overlap, that is once
on each Thin rhombus of the binary tiling.
In a decagonal tiling,
there is an average of $3.19$ Ni on each \Star~
and the composition is $Al_{70}$Co$_{21.2}$Ni$_{8.8}$ 
(assuming the TM fraction is 30\%). (One other version of
the idealization, including the placement of Al atoms, 
is described by Gu~\et~\cite{Guet2005b}, and gave
a composition Al$_{70.1}$Co$_{22.4}$Ni$_{7.5}$, and
it is even easier to devise variants with higher Ni content.)

This model does not include the `pentagonal bipyramid'
motif that is prominent in the related approximant W(Al-Co-Ni)
\cite{Mihalkovic2006}, so we have not 
addressed whether that motif is favoured in the decagonal
Al-Co-Ni structure~\cite{Deloudi2006}.

\section{Orientational order and puckering}

A full structure description must specify,
in each \Decagon's first ring,
the orientation of the pentagon  formed by its ring 1
TM atoms (this is prominent in electron microscope images).
Each orientation is parametrized by a  
spin-like variable $\sigma=\pm 1$, 
with $\sigma=+1$ when the Co$_5$ pentagon 
(projected on the $xy$ plane) points up, that iswhen the five Co
lie in the even-numbered 
atom layers.  We seek an {\it effective interaction}
of neighboring \Dec's, like the spin interaction in an Ising model.  
A cluster arrangement with all clusters oriented the
same will be called `ferromagnetic'; that quasicrystal
has global {\it pentagonal} (not decagonal) symmetry, as is
observed in fact for one of the Co-rich modifications
\cite{Ritschet1996, Liet1996}.

\subsection{Fixed site list: interaction via \Stars}

When we simulate with a fixed site list, there are small 
energy differences that favour either the `ferromagnetic'
or `antiferromagnetic' order of \Decagon~orientations,
depending on the atom density assumed and which approximant
is studied, showing there are competing contributions
to the cluster-cluster effective interaction.
(The `ferromagnetic' effect wins for 
0.068 $-$ 0.074 \AA$^{-3}$, which includes
the physically more reasonable densities.)
 
Where does the interaction come from?
The only atoms that change with a cluster's orientation 
are the central Al and the ring 1 Al$_5$Co$_5$,
but those are too distant to interact {\it directly}. 
(Our potentials were cut off around 7\AA{}.)
An {\it indirect} interaction may be
mediated by other atoms in two ways.
Firstly, the variable Al atoms in rings 2.5 and 3 are within
range of ring 1 of both clusters; this guides whether
the ring 3 Al (in projection) divide the \Decagon~ edge in 
the ratio $\tau^{-1}:\tau^{-2}$ or $\tau^{-2}:\tau^{-1}$
(See figure~\ref{fig:clusters}a.)
This contribution favours `ferromagnetic' order, 
so the shared \Decagon~ edges will match. 
(Expressed in energy terms, the argument is related to the one 
for the `puckering' case in Sec.~\ref{sec:pucker}, below.)

Alternatively, neighboring TM atoms in the \Stars~ interact with 
each other and also respond to the first-ring
orientations of adjoining \Decagons: we suggest this is the
origin of the `antiferromagnetic' coupling, based on
the TM rules of Sec.~\ref{sec:deco}.
Let us assume a repulsive energy $V_{72}$ for
Ni-Ni pairs on edges differing by $72^\circ$ as
found by \Mihet~\cite{Mihet2002} in `basic' Ni modification.
One such pair 
appears on each Fat rhombus having 
\Decagons~ with the same orientation. The effective
orientation interaction is then $\frac{1}{2} (1+\sigma_1\sigma_2) V_{72}$
favouring the `antiferromagnetic' arrangement. Thus, the two
indirect effects are indeed competing.


\section {Relaxation and puckering}
\label{sec:pucker}

Relaxation and molecular dynamics of the structures found in the 
fixed-site simulation were used to explore subtler aspects 
corresponding to small energy differences -- including
the relative orientation of neighboring clusters, which is
a major focus of this paper.
The most striking effect of letting atoms off of the fixed sites
was that Al atoms in ring 2.5/ring 3 tended to run to new locations,
in which they displaced from the layers in the $z$ direction 
(puckering of atom layer).
When we used a simulation cell with four layers, pattern of 
puckering is such that lattice constant doubled to $2c \approx 8$\AA{}. 
(That is the observed period of most Co-rich decagonal Al-Co-Ni modifications and
approximants, and it was long known crystallographically that puckerings
are the major way in which the real cells are doubled.)

\subsection {Channels}
\label{sec:channels}

\begin{figure}
\includegraphics[width=5.0in,angle=0]{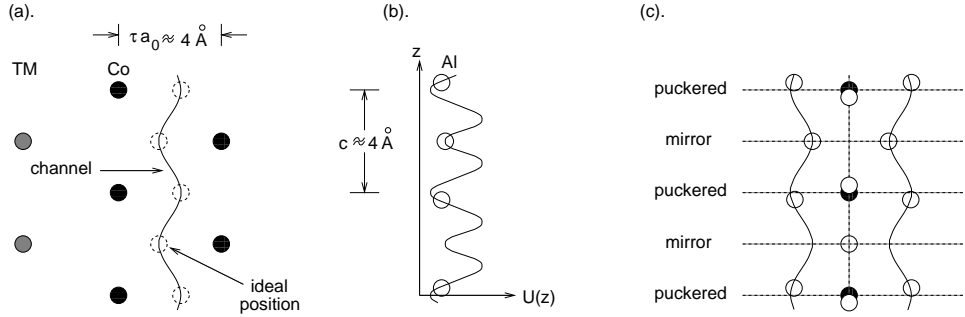}
\caption{`Channels' accomodating three Al atoms per four layers.
(a). Path of a channel favourable to Al between two 
staggered columns of Co atoms; positions of the rigid site
list shown as dashed circles.  The farther TMs (grey) actually
lie in a direction nearly normal to the paper.
(b). Single-Al potential $U(z)$ in the channel, showing 
minimum-energy configuration of three atoms with period $2c$.
(c). Four channels adjacent to single Co column; the 
two channels in the plane normal to the paper break the
period $c$ symmetry in the opposite way from the other two.}
\label{fig:channel}
\end{figure}

How shall we best understand the positions taken by Al after 
they are freed from the fixed site list?
Let us consider the net potential felt by an Al given 
the TM atoms, which we fix on ideal sites 
(they hardly displace under relaxation in any case).
Winding one-dimensional `channels' are found along which the potential is
low and comparatively flat,  typically between 
two columns of Co atoms (a ubiquitous pattern in the structure)
forming a  zigzag ladder, 
as shown in figure~\ref{fig:channel}a.
(\Henet~\cite{Henet2002} discovered similar channels 
in `basic' Ni-rich modifications from the time-averaged Al density in a molecular
dynamics simulation.)

\def\tU{{\tilde U}}

The single-Al potential in a channel may be approximated as 
  \begin{equation}
    U(z) = U_0 - \tU_{c/2} \cos 
       \Bigl(\frac{4 \pi z}{c}\Bigr)  
     + \sigma \tU_{c} \cos 
       \Bigl(\frac{2 \pi z}{c}\Bigr)  
  \label{eq:Uz}
  \end{equation}
Using our potentials, 
$\tU_c\approx \tU_{c/2} \approx 0.2$ eV is found.
Let us explain the origin of these coefficients.
The Co atoms {\it adjacent} to the channel have symmetry under 
one-layer shifts, which explains the period-$c/2$ term of (\ref{eq:Uz}).
Its minimum occurs in each atom layer, where the Al atom
would be equidistant from three Co atoms at $r\approx 2.5$\AA{},
the minimum of the $V_{\rm AlCo}(r)$ pair potential.

Figure~\ref{fig:channel} shows another column of TM farther away
which breaks the degeneracy between even and odd layers.
This explains the period-$c$ term in equation (\ref{eq:Uz}), 
in which $\sigma$ is the `spin' variable defined above, 
and $\sigma = +1$ (resp. $-1$)  
when the distant Co atoms are in even (or odd respectively) layers.
(The deeper minima occur in the TM-poor layer.)
Note that $U_{c}$ is proportional to 
$-d V_{\rm AlTM}(r)/dr$,  which happens to have a maximum 
at $r\approx 4 $\AA{}, the distance to the distant column.


\subsection {Puckering}

Now, {\it three} Al atoms can and do fit into every four layers:
their mean separation $2c/3 \approx 2.7$\AA{} is right at
the Al-Al hardcore radius.
Just where do they prefer to lie in each channel?
The optimum configuration has a symmetry such that
$z_{3m}= m(2c)$, $z_{3m\pm 1} = z_{3m}\pm (2c/3+u)$, 
[so the puckering displacement from a layer is $c/6+u$;
here $m$ is any integer.]
The total energy (per Al) is 
    \begin{equation}
        E \equiv \frac{1}{3}\sum _{m=-1}^{+1} [U(z_m) + V(z_{m+1}-z_m) ], 
    \label{eq:Etot}
    \end{equation}
where $V(z'-z'')$ is the Al interaction along the channel.
[Note a small abuse in this approximation: since the channels
are not straight, the actual Al-Al distance is not just a 
function of $z$~\cite{Guet2005b}.
Inserting $z_m$ into equation (\ref{eq:Etot}) and Taylor expanding in $u$,
 we obtain
    \begin{equation}
       E = E_0 + \frac{2}{3} {U_1}' u + \Biggl( V_1'' + \frac{1}{3} {U_1}'' \Biggr)
              u^2 .
    \label{eq:Eu}
    \end{equation}
Here $E_0 \equiv U_0+V(2c/3)$. Also, ${U_1}' \equiv U'(2c/3)$, etc; 
inserting their values from
equation (\ref{eq:Uz}), 
minimizing with respect to $u$, and using ${U_1}'' \ll {V_1}''$,
we find the energy minimumn
    \begin{equation}
       E_{\min} = E_0 - \Bigl(\frac{\pi}{c}\Bigr)^2 
        \frac{(2 U_{c/2} + \sigma U_c)^2 }{3V_1''}.
    \label{eq:Emin}
    \end{equation}
It can be seen this solution is favourable when $\sigma >0$; that is,
the {\it local} mirror layer of the Al's in each channel 
lies in the TM-rich layer [as observed in real Al-TM structures.]
Then, when two (equi)distant TM columns affect the channel, 
$\sigma \to \sigma_1+\sigma_2$ in
equation (\ref{eq:Emin}) and the lowest $E_{\rm min}$ is
obtained when $|\sigma_1+\sigma_2|=2$. 
Thus the relaxed total 
energy of channel Al favours a `ferromagnetic' alignment
of orientations, as was claimed.

Each channel breaks $z$-translation symmetry in that 
the mirror-layer Al sits at either $z=0$ or $z=c$, but not both.
These channels interact, in a pattern~\cite{Guet2005b}
whereby {\it four} channels get occupied around 
a single TM column, as shown schematically 
in figure~\ref{fig:channel}c; viewed
in projection, the mirror-layer Al atoms involved in
the channels form a sort of bent cross.


({\sl Note added.}
The puckering instability is presumably
related to some very low frequency phonon modes~\cite{Hiramatsu2006}; 
the channels are also the conduit for the 
strong Al diffusion~\cite{Hockeret2006}.)

\section{Conclusion}

To summarize, the methods developed by \Mihet~\cite{Mihet2002} in a study of
the `basic' Ni-rich decagonal Al$_{70}$Co$_{10}$Ni$_{20}$ quasicrystal
has successfully
led to a structural understanding (at $T=0$) of 
`basic' Co-rich
decagonal Al$_{70}$Co$_{20}$Ni$_{10}$, which is described by 
a completely different tiling, despite 
a very similar nearest-neighbor order.
In this case, the key motif is a $12.6 $\AA{}~
diameter decagon having a pentagonal core. 

We applied our approach to a unit cell the same size as
W(Al-Co-Ni), a crystalline approximant of `basic Co'.
Many details of our prediction (most importantly the \Decagons,
and the locations of their centres)  are in agreement with
the experimentally solved structure~\cite{Sugiyamaet2002}.
However, W(Al-Co-Ni) has a few highly puckered sites which 
can not be captured in our approach which begins using
a fixed-site list and relaxes afterwards.
[Mihalkovi\v{c} and Widom~\cite{Mihalkovic2006} have given a 
more thorough ab-initio investigation of W(Al-Co-Ni).]

In general, the simulations with a fixed-site list -- followed
blindly -- would lead to an imperfect description of Al atoms in 
ring 2.5 and ring 3 of the \Decagon.  In simulations 
with relaxations (and in reality), many of
these atoms displace from the layers, in a fashion which
we have explained on the basis of the pair potentials.

\ack 
This work was supported by DOE grant
DE-FG02-89ER-45405.

\end{document}